\documentclass[aps,prl,superscriptaddress,twocolumn]{revtex4}
\usepackage{graphics}
\usepackage{graphicx}
\usepackage{color}
 \usepackage{amsmath, amsthm, amssymb}
 \usepackage{subfigure}

\newcommand{\ep}{\epsilon}

\begin{document}

\title{Non-Equilibrium Transport Through a Gate-Controlled Barrier on the Quantum Spin Hall Edge  }%

\author{Roni Ilan}
\email{ rilan@berkeley.edu}
\affiliation{Department of Physics, University of California, Berkeley, California 94720, USA}

\author{J\'er\^ome Cayssol}
\affiliation{Department of Physics, University of California, Berkeley, California 94720, USA}
\affiliation{Max-Planck-Institut f\"ur Physik Komplexer Systeme, N\"othnitzer Str. 38, 01187 Dresden, Germany}
\affiliation{LOMA (UMR-5798), CNRS and University Bordeaux 1, F-33045 Talence, France}

\author{Jens H. Bardarson}
\affiliation{Department of Physics, University of California, Berkeley, California 94720, USA}
\affiliation{Materials Sciences Division, Lawrence Berkeley National Laboratory, Berkeley, CA 94720}

\author{Joel E. Moore}
\affiliation{Department of Physics, University of California, Berkeley, California 94720, USA}
\affiliation{Materials Sciences Division, Lawrence Berkeley National Laboratory, Berkeley, CA 94720}

%\date{June 2012}

\begin{abstract}
The Quantum Spin Hall insulator is characterized by the presence of gapless helical edge states where the spin of the charge carriers is locked to their direction of motion. In order to probe the properties of the edge modes, we propose a design of a tunable quantum impurity realized by a local gate under an external magnetic field. Using the integrability of the impurity model, the conductance is computed for arbitrary interactions, temperatures and voltages, including the effect of Fermi liquid leads. The result can be used to infer the strength of interactions from transport experiments.
\end{abstract}

%\pacs{}

\maketitle

The quantum spin Hall effect (QSHE) is a property of certain two-dimensional electron systems with strong spin-orbit coupling~\cite{Hasan:2010ku,Qi:2011hb}.  The bulk of the system is electrically insulating, while a conducting ``helical edge'' exists at the boundary in which  electrons of opposite spin move in opposite directions~\cite{Kane:2005hl,Kane:2005gb,Bernevig:2006ij}. Due to this reduction of the number of degrees of freedom, the QSHE edge is expected to realize the physics of a spinless Luttinger liquid, as opposed to a conventional one-dimensional wire that represents a spinful Luttinger liquid~\cite{Giamarchi:2004uc}. The Luttinger liquid is the generic state of metallic interacting electrons in one dimension~\cite{Haldane:2000bk}, while metallic electrons in higher dimensions typically form a Fermi liquid.

The QSHE is realized in (Hg,Cd)Te quantum wells~\cite{Konig:2007hs,Roth:2009bg} where measurements of the conductance indicate the existence of helical edge modes. The simplest measurement to perform on such a system would be a two-terminal conductance measurement. Such a measurement can confirm that the current is carried by helical one-dimensional edge channels, but it can neither provide information on the interaction strength within those channels, nor verify the expected Luttinger liquid behavior. This is the case because, when a clean interacting wire is placed between Fermi liquid contacts (modeled as a non-interacting wires), the measured conductance is insensitive to the interactions~\cite{Maslov:1995je,Safi:1995br}.  

As it turns out, there is a way in which the two terminal conductance can provide information on the interaction strength within the edge modes. A common way of studying one-dimensional systems, both theoretically and experimentally, is by exploring impurity effects on measurable quantities such as their conductance. In general, the problem becomes quite involved when interactions are present, and one usually has to rely on the asymptotic behavior of such quantities (at high or low temperatures, for example) to extract information on the interaction strength. However, in some unique cases certain properties of the edge model make it possible to obtain exact solutions. The QSHE edge is an example of such a system, since the model of a spinless Luttinger liquid with an impurity is ``integrable" ~\cite{Fendley:1995im}. 

In order to utilize the powerful tool of integrability to describe actual measurements on a QSHE edge, backscattering must be induced within a single edge (the model describing backscattering between the two edges of the QSHE system is not integrable). In principle, this can be done by means of a magnetic impurity that locally breaks time-reversal symmetry. However, it is much more desirable to find a way to engineer an impurity with a tunable strength, in order to induce the crossover between weak and strong backscattering. 

In this work we consider combining the effects of an externally applied magnetic field and a local gate voltage to form an artificial impurity on the QSHE edge. The magnetic field direction is carefully chosen such that it breaks time-reversal symmetry yet leaves the edge modes gapless. These edge modes, now unprotected, become sensitive to the local perturbation generated by the gate in the form of an induced Rashba spin-orbit coupling. The strength of the impurity is set both by the magnetic field and the gate voltage. With controlled means for introducing an impurity, the integrality of the edge model~\cite{Saleur:1998vm,Fendley:1995im,Fendley:1995cs} allows us to extract the shape of the non-equilibrium, finite temperature conductance curve, which strongly depends on the value of the Luttinger parameter. Hence measuring the conductance throughout the crossover from weak to strong backscattering could provide information on the interaction strength within the edge channels.

\begin{figure}[t!]
\includegraphics[height=2.2in]{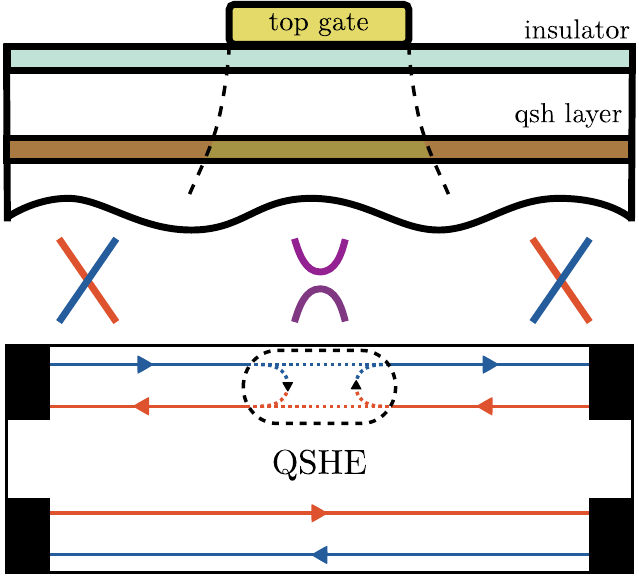}
\caption{Schematic diagram of the proposed experimental setup. Voltage the top gate is used to locally tune the strength of the Rashba spin-orbit coupling. Combined with a magnetic field aligned along the electron spin quantization axis, a gap appears in the edge spectrum, giving rise to backscattering in the helical edge.}
\label{fig:setup}
\end{figure}

The setup we have in mind (see Fig.~\ref{fig:setup}) is similar in spirit to a quantum point contact in fractional quantum Hall effect (FQHE) devices~\cite{Chang:2003bx}. There, particle backscattering between modes with opposite chirality is enhanced with the aid of two gates depleting the electron density and bringing the two edges of the sample closer together. However, for the QSHE device we consider, backscattering between counter-propagating modes takes place on the same edge. Hence, we do not require that the two edges of the sample be brought together, and a single gate is sufficient. Recently, leading corrections to the linear conductance induced by a generic magnetic impurity in a fractional topological insulator were calculated~\cite{Beri:2012jx}. There, unlike the integer case we study, an edge with repulsive interaction can be stable to magnetic perturbations. 

Note that although both the QSHE and the FQHE edges realize a spinless Luttinger liquid, the Luttinger parameter for the QSHE can in principle obtain any value, while for the FQHE it is restricted to quantized values. Another crucial difference between the two systems is embodied in the effect of Fermi liquid contacts discussed earlier. For the FQHE, contacts are expected to have no effect on the conductance, due to the spatial separation of modes of opposite chirality. This has been observed in experiments~\cite{Milliken:1996gp,Du:2009ce,Bolotin:2009ko}. Therefore, the QSHE case has the potential to provide the first experimental test of integrability at non-quantized values of the Luttinger parameter and in the process verify the effects associated with Fermi-liquid contacts.

We start by considering the non-interacting case, solving the scattering problem of two gapless regions separated by a finite strip in which an energy gap is present. We find the reflection strength and show that it can display resonant behavior for some values of the parameters. We then consider interactions and use a method known as the thermodynamic Bethe ansatz to obtain the non-equilibrium finite temperature conductance for various values of the Luttinger parameter \cite{Saleur:1998vm}.

The low energy physics of the non-interacting edge in the presence of a magnetic field $B$ and a position dependent Rashba spin-orbit coupling $\alpha(x)$ is described by the Hamiltonian
\begin{equation}\label{NIH}
H=-i\hbar v_F\sigma_z\partial_x+\mu_B\frac{g_e}{2}\vec{B}\cdot\vec{\sigma}-\frac{i\hbar}{2}\{\alpha,\partial_x\}\sigma_y,
\end{equation} 
where $v_F$ is the Fermi velocity, the $\sigma$'s are the Pauli matrices, $\{.,.\}$ denotes an anticommutator, $g_e$ is the electron Land\'e $g$-factor and $\mu_B$ is the Bohr magneton. To simplify notation, in the following we take $\hbar=1$ and define $M=\mu_Bg_eB/2$. For $M=\alpha=0$ the spectrum of this Hamiltonian is  gapless, $E=\pm v_F p $. When a magnetic field is turned on, the energy spectrum becomes gapped, unless the magnetic field is parallel to the spin quantization axis of the electron.  In that case the effect of the field is merely to shift the Dirac point and $E=\pm (v_F p + M)$. In the absence of a magnetic field, a finite constant spin-orbit interaction $ \alpha(x)=\alpha_ 0$ renormalizes the electron velocity to $v_\alpha=\sqrt{\alpha_0^2+v_F^2}$,  and rotates the electron spin quantization axis by an angle $\cos\theta=v_F/v_\alpha$ about the $x$ axis~\cite{Vayrynen:2011hr}. Note that the spins of the counter-propagating modes remain anti-parallel in the presence of the Rashba term as required by time-reversal symmetry.  

Let us now consider a system in which the magnetic field is uniform and points along the spin quantization axis,  while a finite constant Rashba coupling exists only within a finite strip of width $d$, $\alpha(x)=\alpha_0\Theta(x)\Theta(d-x)$. Outside the strip ($x<0, x>d$), the energy spectrum is gapless, while within the strip ($0<x<d$), the external magnetic field is no longer aligned with the spin polarization axis, and the energy spectrum becomes gapped 
\begin{equation}\label{spectrum}
E=\pm\sqrt{(v_\alpha^2 p + v_F M/v_\alpha)^2  + \alpha_0^2 M^2/v_\alpha^2} ,
\end{equation}
with the energy gap $E_g=|2 \alpha_0 M/ v_\alpha|$. 
In the presence of the external field, the two otherwise decoupled spinors now mix in the region combining both the field and the spin-orbit coupling. The result is a square scattering barrier, from which incoming waves can be reflected. In the limit of a narrow constriction, this region acts as a localized impurity in our helical quantum wire, whose strength is controlled by $M$ and $\alpha_0$. In reality this can be realized by varying the voltage of a nearby electrostatic gate which enhances the Rashba coupling in the vicinity of the gate, while the Rashba coupling far from the gate is negligible~\cite{Nitta:1997gk}. 

We solve the scattering problem by defining  the scattering state in each region to be
\begin{eqnarray*}
   \Psi(x)=\begin{cases}
      \psi_{R}e^{ip_{R}x}+ r\psi_{L}e^{ip_{L}x} &x<0\\ 
      a_+\psi_+e^{ip_+x}+a_-\psi_-e^{ip_-x} & 0<x<d \\ 
      t\psi_{L}e^{ip_{L}x} & x>d \\ 
  \end{cases}
 \end{eqnarray*}
where $\psi_{R}=(1,0)$, $\psi_{L}=(0,1)$ and $\psi_\pm=(i\alpha p_\pm,v_F p_\pm -M-E)$. The momenta $p_{R/L}=(\pm E -M)/v_F$ correspond to the right (R) and left (L) movers outside the strip, while $p_\pm$ are the two momenta inside the strip, corresponding to the solutions of~\eqref{spectrum} at a given energy. The nontrivial part of the solution  for $r$ and $t$, the reflection and transmission amplitudes is to find the correct matching condition for the wave function $\Psi$ at $x=0,d$.

For a general profile  of $\alpha(x)$,  the Schroedinger equation $H\psi=E\psi$ [Eq.~\eqref{NIH}] can be solved formally as $\Psi(x_1) = T_{x_1,x_0} \Psi(x_0)$ where the transfer matrix is written as
\begin{equation}\label{FORMAL}
	T_{x_1,x_0} = P_{x}e^{i\int_{x_0}^{x_1} dx \frac{ (v_F \sigma_z + \alpha \sigma_y)}{v_F^2+\alpha^2} \left[E+M\sigma_z + \frac{i}{2} (\partial_x \alpha) \sigma_y \right]},
\end{equation}
with $P_x$ representing the path ordering operator. For a step in $\alpha$,  $\alpha(x)=\alpha_0\Theta(x)$, we set $x_0=-\delta$ and $x_1=\delta$ and then take the limit of $\delta\rightarrow 0$. The contribution of the terms including the magnetic field and the energy in the exponent will vanish, and we are left with the matching condition
\begin{equation}\label{BC}
\psi(0^+) =\left(  \frac{v_F}{v_\alpha}  \right)^{1/2} e^{i\theta_0\sigma_x}\psi(0^-) ,
\end{equation}
with $\tan2\theta_0=\alpha_0/ v_F$. 

\begin{figure}[tb!]
\begin{center}
\includegraphics[width=0.98\columnwidth]{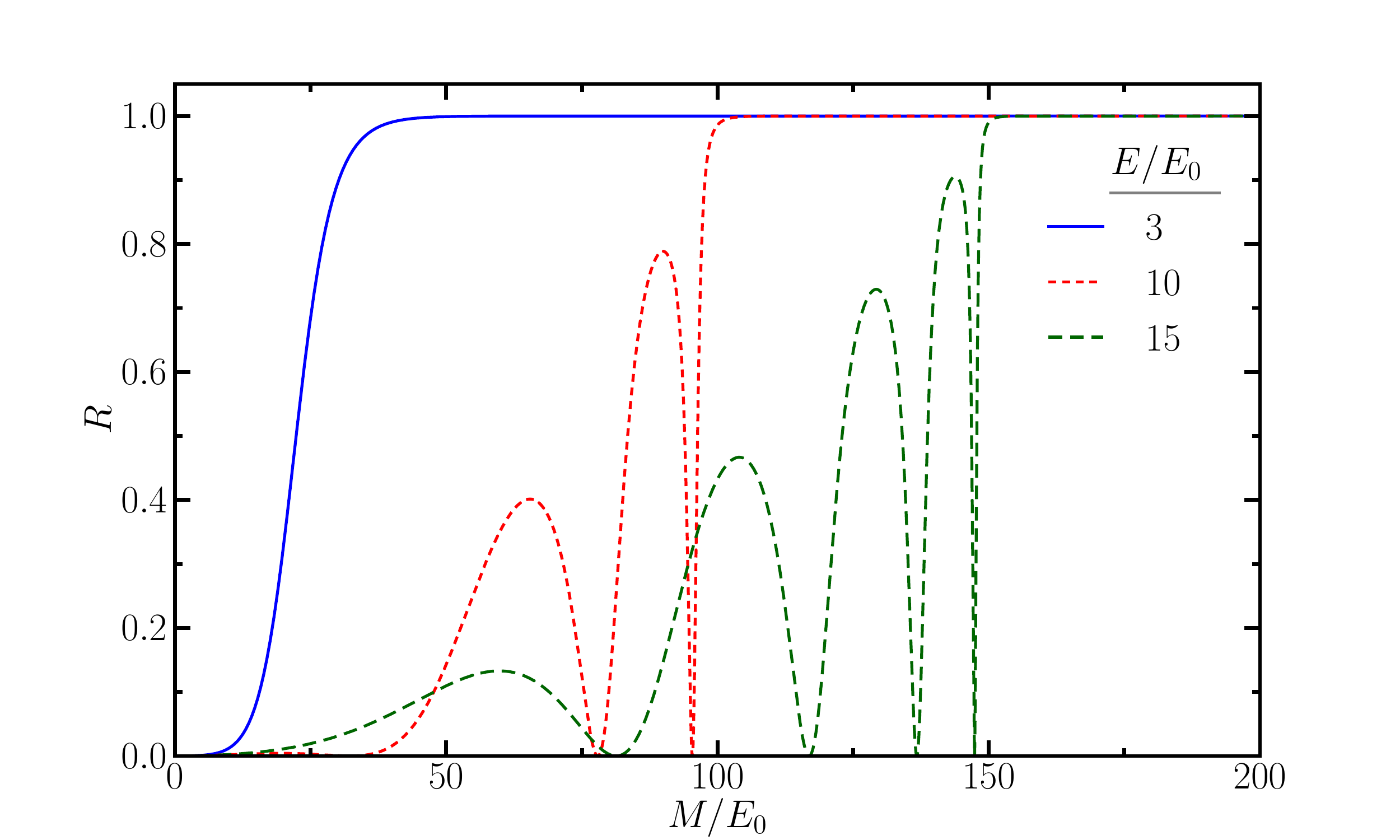}
\caption{Reflection probability $R$ as a function of the normalized magnetic field $M$ for various Fermi energies $E$, and $\alpha_0/v_F=0.1$. The energy unit is $E_0=\hbar v_F / d$.}
\label{fig:reflection}
\end{center}
\end{figure}

Using the boundary condition~\eqref{BC} at $x=0$, and a similar one at $x=d$, we obtain the solutions for $r$, $t$. The analytical form of these solutions is lengthy, therefore we will not present it here but rather plot the reflection probability $R=|r|^2$ in Fig.~\ref{fig:reflection}. The behavior of $R$ as a function of the field at a fixed (nonzero) value of $\alpha_0/v_F$ can be described as follows: at zero field, the reflection is zero, while for large fields $M\gg v_\alpha E/\alpha_0$, the reflection is perfect, $R=1$. In between, $R$ can display two types of behaviors, depending on whether the waves inside the barrier are evanescent or propagating. For evanescent waves $E^2<\alpha_0^2M^2/v_\alpha^2$, $R$ rises monotonically towards unity, while propagating waves result in an oscillating reflection amplitude, due to Fabry-P\'erot type of interference resonances.  The condition for a resonance is simply $(p_+-p_-)d=2\pi n$, and depends both on the value of $\alpha_0$ and $M$. Such Fabry-P\'erot resonances have also been predicted for a QSHE edge state under a spatially inhomogeneous magnetic field \cite{Soori:2011vi}. 

While the resonances appearing for the finite barrier would provide a test for the existence of helical edge modes in the absence of interactions, it is expected that interaction effects are important in $1$D systems and can renormalize drastically the backscattering created by a single impurity~\cite{Kane:1992kx}. This is true even for weak repulsive interactions since single electron backscattering is described by a relevant operator (in the renormalization group sense), leading to a crossover from a weak-backscattering regime to a strong backscattering regime as the temperature is lowered. The non-interacting solution is still useful in estimating the bare backscattering strength and its dependence on $\alpha_0$ and $M$. In our discussion of the non-interacting problem we considered a region of finite width $d$ as the scatterer. In order to calculate conductance in the interacting case we need to consider a point like scatterer. The bare backscattering strength for such an impurity could be estimated from our previous calculation by taking the limit of a very narrow barrier \footnote{Strictly speaking, in the limit of zero width the transmission for this problem becomes unity.}.  For weak magnetic fields and small $\alpha_0$ it is simply given by $R\sim (M\alpha_{2k}/v_F^2)^2$, where $\alpha_{2k}$ is the $2k$ component of the Fourier decomposition of $\alpha(x)$.

The Hamiltonian of the interacting QSHE edge in the bosonization language is  
\begin{equation}
H=\frac{v}{4\pi g}\int dx (\partial_x\phi_R)^2+(\partial_x\phi_L)^2 ,
\end{equation}
where $\phi_{R/L}$ are left moving and right moving boson fields,  $g$ is the Luttinger liquid parameter and $v$ is the edge velocity renormalized by interactions~\cite{Teo:2009bf}. A backscattering term couples to $\phi_L-\phi_R$
\begin{equation}
H_B=\lambda\cos(\phi_L(0)-\phi_R(0)) .
\end{equation} 
By defining even and odd non-local combinations of the fields $\phi_{\text{e}/\text{o}}=1/\sqrt{2}(\phi_L(x,t)\pm\phi_{R}(-x,t))$, the backscattering term couples only to $\phi_\text{o}$, and the Hamiltonian breaks into two decoupled contributions. The part of the Hamiltonian describing the odd fields is integrable, since it is identical to the massless limit of the boundary sine-Gordon (SG) model~\cite{Fendley:1995im,Fendley:1995cs}.  The even field theory is free and does not interact with the impurity. 

The integrability of the SG model was previously used by Fendley, Ludwig, and Saleur (FLS)~\cite{Fendley:1995im,Fendley:1995cs} to calculate the non-linear conductance in a point contact geometry for fractional quantum Hall states at $\nu=1/m$.  In Ref.~\onlinecite{Fendley:1996ca} a similar formula is derived for the conductance at $\nu=1-1/m$ by exploiting the relation between the Kondo problem in a magnetic field and the SG model at finite voltage.  Here, we use the same method to compute the differential conductance curves at $m=3,4,5$ in order to demonstrate the behavior of the QSHE edge transport at $1/2<g<1$. A result for continuously variable $g$ can in principle be computed using a more involved technique developed in~\cite{Dorey:1999da}. Also, if the repulsive interactions are weak ($g$ close to $1$), a more direct approach based on resummed perturbation theory~\cite{Yue:1994ha} can be used, but it is currently unclear whether interactions are weak in the QSHE edge.

For $g=1-1/m$, the current along the QSHE edge is given by~\cite{Fendley:1996ca}: 
 \begin{align}
\label{G}
&I(V,T_B,T)=\frac{T(m-1)}{2} \int   \frac{d\theta}{\cosh^2[\theta-\ln(T_B/T)]}\times  \\ &\ln \left(  \frac{1+e^{(m-1)V/2T-\ep_{+}(\theta)}  }{1+e^{-(m-1)V/2T-\ep_{+}(\theta)} }  \frac{1+e^{-(m-1)V/2T-\ep_{+}(\infty)} }{1+e^{(m-1)V/2T-\ep_{+}(\infty)}  } \right).\notag
\end{align}
Here  $T_B$ is an energy scale related to the impurity strength $\lambda$ by $T_B=C\lambda^{1/(1-g)}$~\cite{Fendley:1995cs,Fendley:1996ca}, where $C$ is a non-universal (cutoff dependent) constant, $\epsilon_{+}$ is the quasi-energy of the kink solution of the sine-Gordon model, and $\theta$ the rapidity.  The energy $\epsilon_+(\theta)$ of the kinks is computed numerically by solving a set of coupled integral equations obtained from the thermodynamic Bethe ansatz. The full details of these equations are given in supplementary material. 

\begin{figure}[tb!]
\includegraphics[width=0.98\columnwidth]{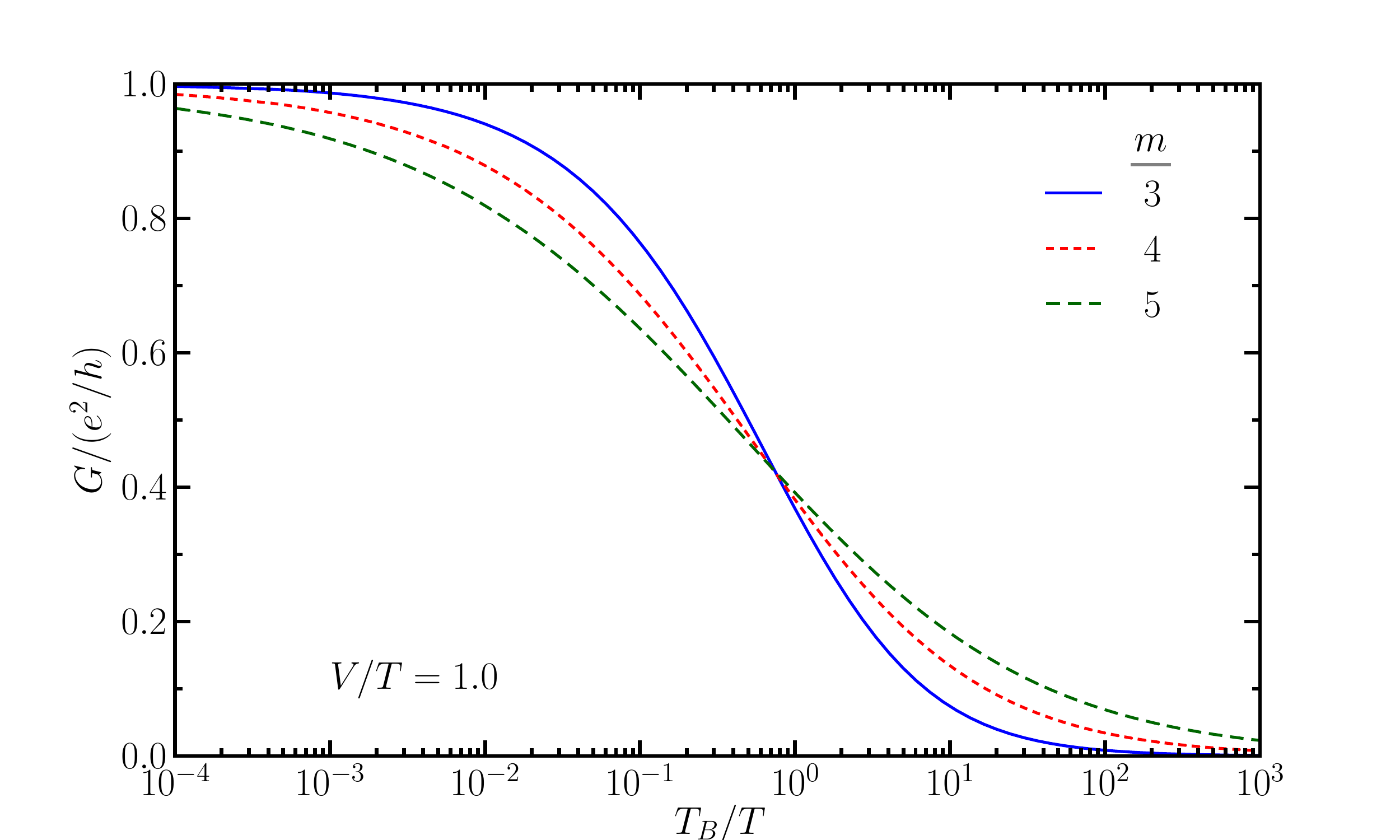}\\
\includegraphics[width=0.98\columnwidth]{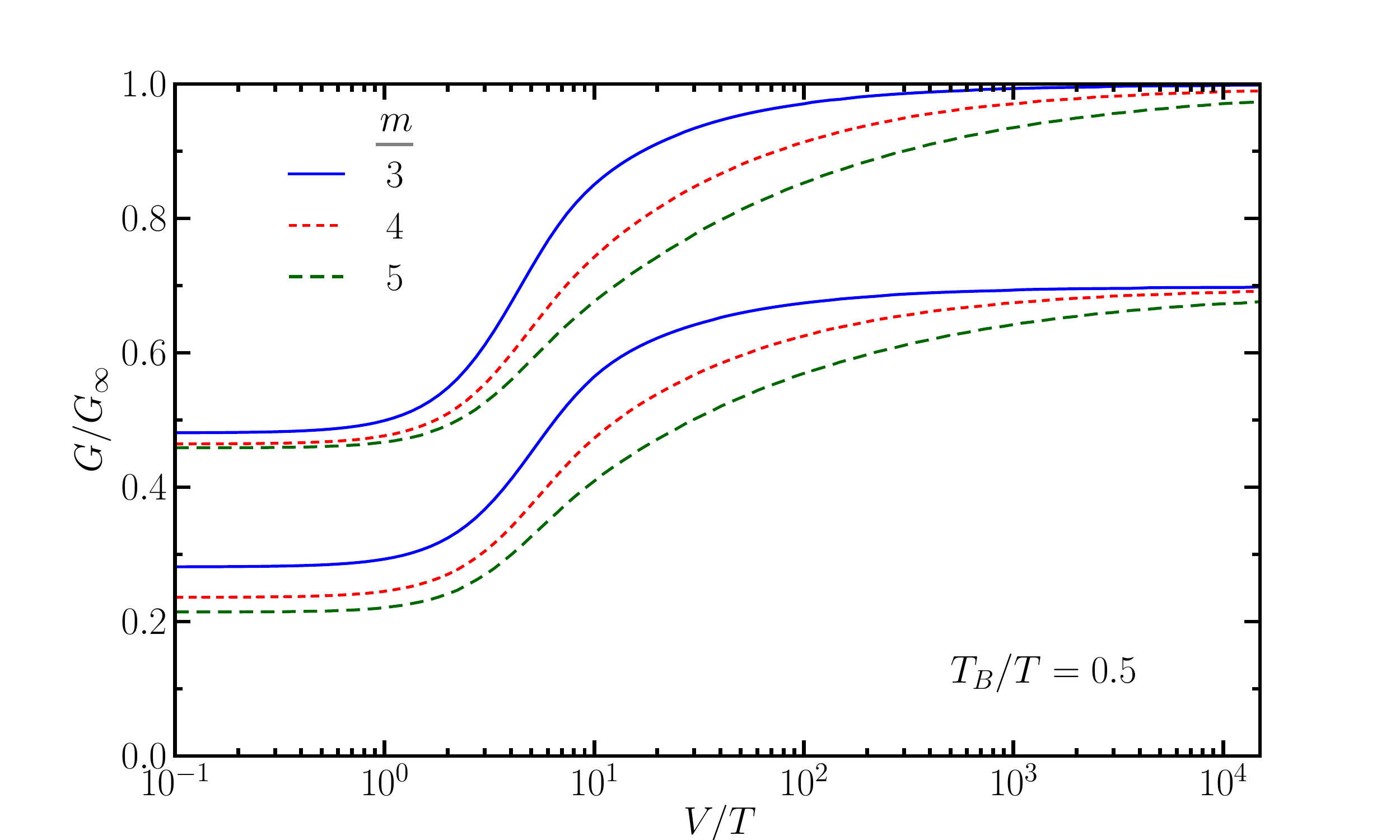}\\
\includegraphics[width=0.98\columnwidth]{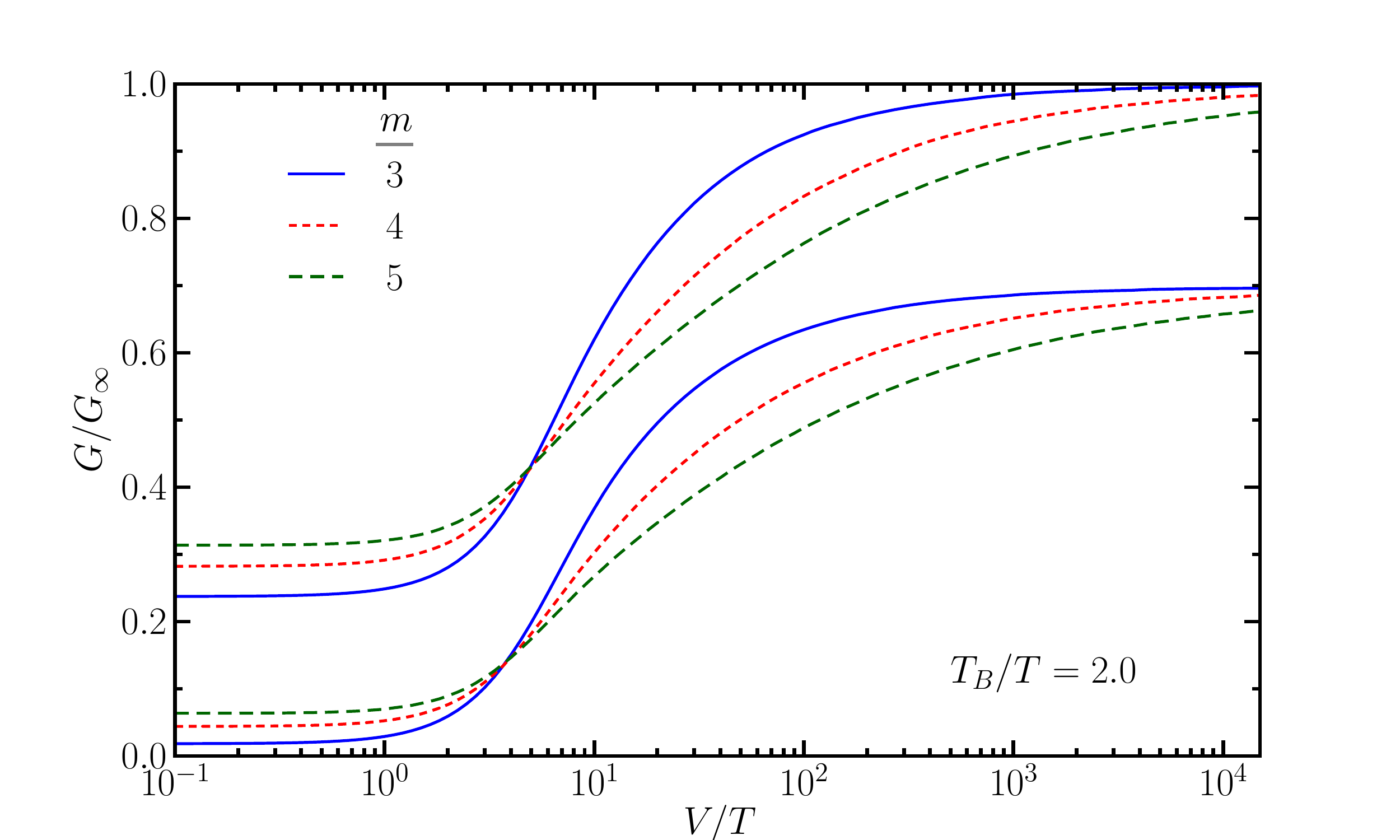}
\caption{Differential conductance for different values of $g=1-1/m$. Top figure: $G=dI/dV$ (with contact corrections) as a function of $T_B/T$. Bottom figures: $G=dI/dV$ as a function of $V/T$ for two different values of $T_B/T$. The differential conductance has been scaled with the value $G_\infty = e^2/h$ and $(1-1/m)e^2/h$ with and without contact corrections, respectively. The curves without contact corrections have been shifted down by $0.3$ for clarity.}
\label{3figs}
\end{figure}

To account for the effect of non-interacting leads on the calculation of the conductance, we adapt a result from Ref.~\onlinecite{Koutouza:2001ki}, where a self-consistency condition was derived for the chemical potential of the various excitations inside the wire, which is not equal to the external applied voltage.  The consequences of this self-consistency condition were extensively explored in Ref.~\onlinecite{Koutouza:2001ki} for $g=1/m$.   Here we carry out a similar analysis for $g=1-1/m$.  Denoting the chemical potential for the kinks and anti-kinks by $\mu_{\pm}=\pm(m-1)W/2$ and the external voltage by $V$, the self-consistency condition for $W$ is~\cite{Koutouza:2001ki}
\begin{equation}
V=- \frac{e^2}{h} \left( 1-\frac{1}{g} \right) I(W) + W .
\label{eq:CC}
\end{equation}
The results for the differential conductance $G=dI/dV$, with and without the contact correction, are presented in Fig.~\ref{3figs}. The asymptotic behavior of $G$ as a function of $T_B/T$ matches the known predictions \cite{Kane:1992kx}, namely $G \simeq e^2/h\left( T/T_B \right)^{2/g-2}$ at low temperature, and $G  \simeq e^2/h\left(1-\left(  T_B/T\right)^{2(1-g)} \right)$ at high temperature.

Though we have computed the full curve for particular values of $g$, all curves show similar features and a comparison with experimental data should confirm the expected Luttinger liquid behavior and yield a good estimate for the value of $g$.  Note that when contact corrections are included the conductance always saturates to $e^2/h$ at high voltage or in the absence of backscattering. Nevertheless, the curve shape itself highly depends on $g$, and in particular, the exponents of the asymptotic behavior remain the same as without the correction.

The feasibility of our proposal depends both on the stability of the QSHE edge in presence of a magnetic field and the spin direction of the modes. The behavior of the QSHE under a magnetic field has been studied in an experiment where the conductance was measured for various tilt angles of the field with respect to the plane of the $2D$ electron gas \cite{Konig:2007hs,Konig:2008bz,Qi:2011hb}. The results show that on top of the contribution from the Zeeman coupling, when the field is perpendicular to the plane, the conductance drops rapidly with the field strength due to orbital effects. Nevertheless, a peak in the conductance of typical width $B=10mT$ exists at $T=30mK$. Orbital effects result in an effective $g$-factor values of $20-50$, the typical Fermi velocity is estimated to be  $v_F = 5.5\times10^{5}m/s$, and $\alpha_0\approx 5\times 10^4 m/s$. Therefore, even under the most restrictive conditions one can obtain a gap size of $E_g \approx 100 - 300 mK$ in the vicinity of the gate in our setup. 

In our analysis we have made the assumption that the spin quantization axis far from the gated region is fixed along the edge, and therefore it is possible to align the magnetic field such that it does not gap out the edge modes in those regions. In principle, the preferred spin quantization axis along the edge is determined by the properties of the material, is not protected, and may tilt in a complicated way along the edge due to fluctuations of the Rashba coupling. However, if the edge is made smooth enough it is reasonable to assume that fluctuations in the Rashba coupling have a much smaller effect on the edge states than the intentional coupling induced by the gate, and therefore our analysis remains valid.   

To summarize, in this paper we proposed a realistic and controlled setup for transport experiments on the QSHE edge using the combined effect of time-reversal symmetry breaking and induced spin-orbit coupling. Using the integrability of the resulting edge model we predict the form of the non-linear conductance, which could be compared with future experimental data. 

We would like to thank Paul Fendley for useful discussions. The authors acknowledge support from AFOSR MURI (RI), the ONR EU/FP7 under contract TEMSSOC and from ANR through project 2010-BLANC-041902 (ISOTOP) (JC), the Nanostructured Thermoelectrics program of LBNL  (JHB), and DARPA (JEM).

\bibliography{qshpc}

\newpage\appendix*
\section{Supplementary material: Thermodynamic Bethe Ansatz equations}

The thermodynamic Bethe ansatz equations relevant for our particular model are the following coupled integral equations~\cite{Fendley:1996ca}:

\begin{eqnarray}\nonumber
&&\ep_j(\theta)/T=\delta_{j1} e^\theta - \sum_{k} N_{jk} \int \frac{d\theta'}{2\pi} \frac{1}{\cosh(\theta-\theta')} L_k(\theta'),\\
&&L_k(\theta')= \ln \left( 1+e^{-(\ep_k(\theta')+\mu_k)/T} \right).
\label{TBA}
\end{eqnarray}
In this set of coupled integral equations, the indices $j$ and $k$ label the  different excitation of the SG model. There are three types of excitations: breathers ($1,..,m-2$), kinks ($+$) and the antikinks ($-$). The functions $\ep_k$ label their energies (parametrized by an angle $\theta$ known as the rapidity). Kinks and anti-kinks are the charge carrying excitations of positive and negative charge respectively, while the breathers are neutral bound states of a kink and an anti-kink, and therefore their chemical potential is alway set to zero.  In general, $\ep_{+}=\ep_{-}$, and for $\nu=1-1/m$, the matrix element $N_{jk}=1$ if the nodes $j$ and $k$ are connected in the following diagram
\begin{figure}[h!]\label{TBAdiagram}\vspace{-.1in}
\begin{center}
\includegraphics[height=.8in]{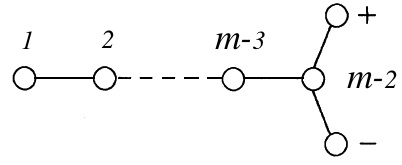}\vspace{-.3in}
\end{center}
\end{figure}

\noindent otherwise $N_{jk}=0$.

We solve these equation numerically for arbitrary values of m. Then the energy
of the kinks is used in Eq.~\eqref{G} to compute the conductance without contact correction, or combined
with the self-consistency condition defined in Eq.\eqref{eq:CC}, to evaluate the conductance in presence of contact corrections.
\end{document}